\documentclass{sample-algotel}
\usepackage[latin1]{inputenc}
\usepackage[francais]{babel}
\usepackage{float}
\usepackage{amsmath}
\usepackage{graphicx}
\usepackage{amssymb}
\usepackage{theorem}
\usepackage{float}
\usepackage{verbatim}

\floatstyle{ruled}
\newfloat{algorithm}{tbp}{loa}
\floatname{algorithm}{Algorithme}

\newtheorem{theo}{Théorème}
\newtheorem{defi}{Définition}
\newtheorem{lem}{Lemme}
\newenvironment{theorem}[1]{\vspace{-0.35cm}\begin{theo}#1}{\end{theo}\vspace{-0.3cm}}
\newenvironment{definition}[1]{\vspace{-0.35cm}\begin{defi}#1}{\end{defi}\vspace{-0.3cm}}
\newenvironment{lemme}[1]{\vspace{-0.35cm}\begin{lem}#1}{\end{lem}\vspace{-0.3cm}}

\author{Swan Dubois\addressmark{1}, Sébastien Tixeuil\addressmark{2} et Nini Zhu\addressmark{3}}

\title{Mariages et Trahisons}

\address{
\addressmark{1} UPMC Sorbonne Universités \& INRIA (France), swan.dubois@lip6.fr\\
\addressmark{2} UPMC Sorbonne Universités \& IUF (France), sebastien.tixeuil@lip6.fr\\
\addressmark{3} UPMC Sorbonne Universités (France)
}


\begin{document}
\maketitle

\begin{abstract} 
Un protocole \emph{auto-stabilisant} est par nature tolérant aux fautes \emph{transitoires} (\emph{i.e.} de durée finie). Ces dernières années ont vu apparaître une nouvelle classe de protocoles qui, en plus d'être auto-stabilisants, tolèrent un nombre limité de fautes \emph{permanentes}. Dans cet article, nous nous intéressons aux protocoles auto-stabilisants tolérant des fautes permanentes très sévères : les fautes \emph{Byzantines}. Nous nous concentrons sur la stabilisation stricte, approche dans laquelle le système doit converger en temps fini vers un comportement tel que l'effet des fautes Byzantines est \emph{confiné} avec un rayon constant autour de chaque processeur fautif (\emph{i.e.} seuls les processeurs corrects situés en dessous d'une certaine distance d'un processeur Byzantin sont autorisés à ne pas respecter leur spécification). Plus spécifiquement, nous étudions la possibilité de construire un mariage maximal dans un réseau où un nombre inconnu et potentiellement non borné de processeurs peut avoir un comportement Byzantin.
\end{abstract}

\vspace{-0.1cm}
\section{Motivations et Définitions}\label{sec:intro}

Le développement des systèmes distribués à large échelle a démontré que la tolérance aux différents types de fautes doit être incluse dans les premières étapes du développement d'un tel système. L'\emph{auto-stabilisation} permet de tolérer des fautes \emph{transitoires} tandis que la tolérance aux fautes traditionnelle permet de masquer l'effet de fautes \emph{permanentes}. Il est alors naturel de s'intéresser à des systèmes qui regrouperaient ces deux formes de tolérance. Cet article s'inscrit dans cette voie de recherche.

Le problème de la construction d'un \emph{mariage maximal} est très étudié en systèmes distribués. \'Etant donné un graphe $G=(V,E)$, un \emph{mariage} $M$ de $G$ est un sous ensemble de $E$ tel que tout sommet de $V$ appartient à au plus une arête de $M$. Un mariage est \emph{maximal} s'il n'existe aucun mariage $M'$ tel que $M\subsetneq M'$. Le problème de la construction d'un mariage maximal est classiquement spécifié de la manière suivante : le protocole termine en un temps fini (vivacité) et dans la configuration terminale, il existe un mariage maximal (sûreté). D'un point de vue réseau, ce problème a de nombreuses applications comme l'allocation distribuée de ressources (fréquences, etc.) dans différents types de réseaux.

\paragraph{Auto-stabilisation} Dans cet article, nous considérons un système distribué asynchrone, \emph{i.e.} un graphe non orienté connexe $G$ où les sommets représentent les processeurs et les arêtes représentent les liens de communication. Deux processeurs $u$ et $v$ sont \emph{voisins} si l'arête $(u,v)$ existe dans $G$. Pour tout processeur $v$, l'ensemble de ses voisins est noté $N_v$. Les variables d'un processeur définissent son \emph{état}. L'ensemble des états des processeurs du système à un instant donné forme la \emph{configuration} du système. Dans cet article, nous nous concentrons sur les problèmes \emph{statiques}, \emph{i.e.} les problèmes dans lesquels le système doit atteindre un état donné et y rester. Par exemple, la construction d'arbre couvrant est un problème statique. De plus, nous considérons des problèmes pouvant être spécifiés de manière locale (\emph{i.e.} il existe, pour chaque processeur $v$, un prédicat $spec(v)$ qui est vrai si et seulement si la configuration est conforme au problème). Les variables apparaissant dans $spec(v)$ sont appelées \emph{variables de sortie} ou \emph{S-variables}.

Un système \emph{auto-stabilisant} \cite{D74j} est un système atteignant en un temps fini une configuration légitime (\emph{i.e.} $spec(v)$ est vraie pour tout $v$) indépendamment de la configuration initiale (propriété de \emph{convergence}). Une fois cette configuration légitime atteinte, tout processeur $v$ vérifie $spec(v)$ pour le restant de l'exécution et, dans le cas d'un problème statique, le système ne modifie plus ses S-variables (propriété de \emph{clôture}). Par définition, un tel système peut tolérer un nombre arbitraire de fautes \emph{transitoires}, \emph{i.e.} de fautes de durée finie (la configuration initiale arbitraire modélisant le résultat de ces fautes). Cependant, la stabilisation du système n'est en général garantie que si tous les processeurs exécutent correctement leur protocole. 

\paragraph{Stabilisation stricte} Si certains processeurs exhibent un comportement \emph{Byzantin} (\emph{i.e.} ont un comportement arbitraire, et donc potentiellement malicieux), ils peuvent perturber le système au point que certains processeurs corrects ne vérifient jamais $spec(v)$. Pour gérer ce type de fautes, \cite{NA02c} définit un protocole \emph{strictement stabilisant} comme un protocole auto-stabilisant tolérant des fautes Byzantines permanentes. Plus précisément, étant donné $c$ (appelé \emph{rayon de confinement}), \cite{NA02c} définit une configuration \emph{$c$-confinée} comme une configuration dans laquelle tout processeur $v$ à une distance supérieure à $c$ de tout processeur Byzantin vérifie $spec(v)$. Un protocole strictement stabilisant est alors défini comme un protocole satisfaisant les propriétés de convergence et de clôture par rapport à l'ensemble des configurations $c$-confinées (et non plus l'ensemble des configurations légitimes comme en auto-stabilisation). Cela permet d'assurer que seuls les processeurs dans le $c$-voisinage (\emph{i.e.} à distance inférieure ou égale à $c$) d'un processeur Byzantin peuvent ne pas vérifier infiniment souvent la spécification.

\paragraph{\'Etat de l'art} Dans le contexte de l'auto-stabilisation, le premier protocole de calcul de mariage maximal a été fourni par Hsu et Huang \cite{HH92j}. Goddard et al.~\cite{GHJS03c} ont ensuite produit une variante auto-stabilisante synchrone de ce protocole. Manne et al.~\cite{MMPT09j} ont finalement fourni un autre protocole pour un environnement asynchrone. En ce qui concerne l'amélioration de la $\frac{1}{2}$-approximation induite par le mariage maximal, Ghosh et al.~\cite{GGHSP95c} et Blair et Manne \cite{BM03c} ont présenté une technique qui peut être utilisée pour calculer un mariage maximum dans un arbre, tandis que Goddard et al.~\cite{GHS06c} ont produit un protocole auto-stabilisant pour calculer une $\frac{2}{3}$-approximation dans les anneaux anonymes de taille non divisible par trois. Manne et al. ont ensuite généralisé ce résultat à toute topologie \cite{MMPT11j}. Notez que, contrairement à notre approche, aucune de ces solutions ne tolère un comportement Byzantin d'une partie du système.

\vspace{-0.1cm}
\section{Mariage Maximal}\label{sec:arbre}

\paragraph{Spécification} Chaque processeur $v$ a une variable $pref_v$ qui appartient à l'ensemble $N_v\cup \{null\}$. Cette variable fait référence au voisin de $v$ qu'il préfère pour un mariage. Par exemple, si $pref_v=u$ alors $v$ veut ajouter l'arête $(v,u)$ au mariage en construction. Pour tout processeur $v$, nous définissons l'ensemble de prédicats suivants : \emph{(i)} $proposition_v$ indique que $v$ propose un mariage à un de ses voisins $u$, mais que $u$ n'a pas encore répondu, \emph{(ii)} $mari\acute e_v$ indique que $v$ a proposé $u$ et que $u$ a proposé $v$ en retour, \emph{(iii)} $condamn\acute e_v$ indique que $v$ a proposé $u$, mais que $u$ a proposé un autre de ses voisins, \emph{(iv)} $mort_v$ indique que $v$ n'a aucun espoir de se marier (chacun de ses voisins est marié à quelqu'un d'autre), et \emph{(v)} $c\acute elibataire_v$ signifie que $v$ n'a proposé personne et qu'au moins un de ses voisins est dans un état similaire. Plus formellement,

\[\begin{array}{ccc}
proposition_v&\equiv&\exists u\in N_v,(pref_v=u)\wedge(pref_u=null)\\
mari\acute e_v&\equiv&\exists u\in N_v,(pref_v=u)\wedge(pref_u=v)\\
condamn\acute e_v&\equiv&\exists u\in N_v,\exists w\in N_u,(pref_v=u)\wedge(pref_u=w)\wedge(w\neq v)\\
mort_v&\equiv&(pref_v=null)\wedge(\forall u\in N_v,mari\acute e_u=true)\\
c\acute elibataire_v&\equiv&(pref_v=null)\wedge(\exists u\in N_v, mari\acute e_u\neq true)
\end{array}\]

Il est trivial de vérifier que pour toute configuration $\gamma$ et pour tout processeur $v$, exactement un de ces prédicats est satisfait par $v$ dans $\gamma$.

Si le système est sujet à des fautes Byzantines, il est évident qu'aucun protocole ne peut satisfaire la spécification classique du problème. \`A partir de maintenant, un processeur $v$ est considéré localement légitime lorsqu'il satisfait le prédicat suivant : $spec(v)\equiv mari\acute e_v\vee mort_v$. Nous pouvons maintenant décrire la propriété globale satisfaite par toute configuration $c$-confinée pour $spec$. Informellement, nous pouvons prouver l'existence d'un mariage maximal sur un sous-ensemble de $G$ qui contient au moins l'ensemble des processeurs $c$-corrects (\emph{i.e.} les processeurs correct à distance strictement supérieure à $c$ de tout processeur Byzantin) dans une telle configuration. Dans la suite, $V_c$ désigne l'ensemble des processeurs $c$-corrects.

\begin{definition}
Pour tout entier $c>0$ et toute configuration $\gamma$, le sous-graphe $G^*_{c,\gamma}$, est le sous-graphe de $G$ induit par l'ensemble de sommets suivant : $V^*_{c,\gamma}=V_c\cup\{v\in V\setminus V_c|\exists u\in V_c,pref_v=u\wedge pref_u=v\}$.
\end{definition}

Il est trivial de déterminer la propriété suivante, satisfaite par toute configuration $c$-confinée pour $spec$.

\begin{lemme}
\label{lemma1}
Dans toute configuration $\gamma$ qui est $c$-confinée pour $spec$, il existe un mariage maximal sur $G^*_{c,\gamma}$.
\end{lemme}

Le résultat du Lemme \ref{lemma1} motive l'écriture d'un protocole strictement stabilisant pour $spec$. En effet, même si la spécification est locale, elle induit une propriété globale dans toute configuration $c$-confinée pour $spec$ étant donné qu'il existe alors un mariage maximal sur un sous-graphe clairement défini.

\paragraph{Modèle} Nous prenons comme modèle de calcul le \emph{modèle à états}. Les variables des processeurs sont partagées : chaque processeur a un accès direct en lecture aux variables de ses voisins. En une \emph{étape} atomique, chaque processeur peut lire son état et ceux de ses voisins et modifier son propre état. Un \emph{protocole} est constitué d'un ensemble de règles de la forme $<garde>\longrightarrow<action>$. La $garde$ est un prédicat sur l'état du processeur et de ses voisins tandis que l'$action$ est une séquence d'instructions modifiant l'état du processeur. A chaque étape, chaque processeur évalue ses gardes. Il est dit \emph{activable} si l'une d'elles est vraie. Il est alors autorisé à exécuter son $action$ correspondante (en cas d'exécution simultanée, tous les processeurs activés prennent en compte l'état du système du début de l'étape). Les \emph{exécutions} du système (séquences d'étapes) sont gérées par un \emph{ordonnanceur} : à chaque étape, il sélectionne au moins un processeur activable pour que celui-ci exécute sa règle. Cet ordonnanceur permet de modéliser l'asynchronisme du système. La seule hypothèse que nous faisons sur l'ordonnancement est qu'il est \emph{localement central} et \emph{fortement équitable}, \emph{i.e.} que deux voisins ne peuvent pas être activés durant la même étape et qu'aucun processeur ne peut être infiniment souvent activable sans être choisi par l'ordonnanceur.

\begin{algorithm}
\caption{$\mathcal{SSMM}$: Mariage maximal strictement stabilisant pour le processeur $v$}
\label{algo1}
\begin{description}
\item[Variables:]~\\
$pref_v\in N_v\cup\{null\}$: voisin préféré de $v$\\
$ancien\_pref_v\in N_v$: voisin préféré précédent de $v$
\item[Fonction:]~\\
Pour tout $u\in\{v,null\}$, $suivant_v(u)$ est le premier voisin de $v$ supérieur à $ancien\_pref_v$ (selon un pré-ordre circulaire) tel que $pref_{suivant_v(u)}=u$
\item [{Règles:}]~\\
/* Mariage */\\
$(M)::(pref_v=null)\wedge(\exists u\in N_v,pref_u=v)\longrightarrow pref_v:=suivant_v(v)$\\
/* Séduction */\\
$(S)::(pref_v=null)\wedge(\forall u\in N_v,pref_u\neq v)\wedge(\exists u\in N_v, pref_u=null)\longrightarrow pref_v:=suivant_v(null)$\\
/* Abandon */\\
$(A)::(pref_v=u)\wedge(pref_u\neq v)\wedge(pref_u\neq null)\longrightarrow ancien\_pref_v:=pref_v;pref_v:=null$
\end{description}
\end{algorithm}

\paragraph{Protocole} Notre protocole de construction de mariage maximal strictement stabilisant, nommé $\mathcal{SSMM}$ est formellement présenté en Algorithme~\ref{algo1}. La base de ce protocole est le protocole de mariage maximal auto-stabilisant de Hsu and Huang~\cite{HH92j}, mais nous autorisons les processeurs à se souvenir de leur précédent  abandon (par exemple, un mariage non respecté par un processeur Byzantin ou une proposition qui n'a pas abouti sur un mariage) dans le but de ne pas répéter les mêmes choix infiniment lorsque des processeurs Byzantins participent au processus global du mariage. Les idées qui sous-tendent le processus de mariage pour les processeurs corrects sont les suivants : \emph{(i)} une fois mariés, les processeurs corrects ne divorcent jamais et ne proposent jamais un autre voisin, \emph{(ii)} les processeurs corrects proposent le mariage à n'importe lequel de leur voisin et acceptent le mariage de n'importe lequel de leur voisin qui le leur propose, \emph{(iii)} si un processeur correct réalise qu'il a proposé le mariage a quelqu'un qui est peut être marié à un autre, alors il annule sa proposition. Un processeur $v$ maintient deux variables : $pref_v$, qui a été présentée dans la spécification du problème, et $ancien\_pref_v$ qui est utilisée pour stocker le dernier voisin à qui $v$ a proposé le mariage de manière non concluante. Enfin, la fonction auxiliaire $suivant_v$ est utilis\'ee pour choisir un nouveau voisin à qui proposer le mariage. Un pré-ordre circulaire est utilisé sur les voisins de façon à limiter l'influence des processeurs Byzantins. En effet, ce pré-ordre circulaire garantit qu'un processeur voisin d'un processeur Byzantin ne le choisira pas immédiatement après un divorce causé par ce Byzantin.

Il est possible de montrer que ce protocole est strictement stabilisant pour $spec$ avec un rayon de confinement de 2. En effet, si on considère une configuration initiale dans laquelle tous les processeurs (même Byzantins) satisfont $spec$ et dans laquelle un Byzantin $b$ est marié avec un voisin correct $v$ et que ce dernier a lui-même un voisin $u$ qui est $mort$, alors $b$ peut forcer $u$ à devenir $c\acute elibataire$ en rompant son mariage avec $v$ (ce qui prouve l'absence de cl\^oture de $\mathcal{SSMM}$ par rapport aux configurations $2$-confin\'ees). De plus, cet exemple est suffisant pour prouver l'optimalité de ce rayon de confinement. On peut donc conclure :

\begin{theorem}
$\mathcal{SSMM}$ est un protocole strictement stabilisant pour la construction d'un mariage maximal avec un rayon de confinement de 2, ce qui est optimal.
\end{theorem}

L'intuition de la preuve de ce r\'esultat est la suivante\footnote{La preuve complète de ce résultat est disponible dans un rapport de recherche \cite{DTZ12r}.}. La cl\^oture de l'ensemble des configurations $2$-confin\'ees est imm\'ediate. La preuve de la convergence du protocole de Hsu et Huang repose sur une fonction de potentiel (\emph{i.e.} une fonction associant une valeur à chaque configuration du système, strictement décroissante pour toute exécution et atteignant une borne inférieure en cas de stabilisation du système, ce qui prouve la convergence en un temps fini du système). Nous avons adapté cette fonction de potentiel (en ne prenant en compte que l'\'etat des processeurs $2$-corrects alors que la fonction originale consid\'erait l'ensemble des processeurs) et montré que cette fonction était ultimement strictement décroissante (\emph{i.e.} les processeurs Byzantins ne peuvent provoquer qu'un nombre fini d'accroissement de la fonction), ce qui est suffisant pour montrer la convergence du protocole en pr\'esence de fautes Byzantines. 

\vspace{-0.1cm}
\section{Conclusion}\label{sec:conclusion}

Dans cet article, nous nous sommes intéressé aux protocoles auto-stabilisants confinant de plus l'effet de fautes Byzantines permanentes. Nous avons montré que le protocole de Hsu et Huang connu pour construire un mariage maximal de manière auto-stabilisante ne nécessitait que très peu de modifications afin de devenir strictement stabilisant, c'est-à-dire qu'il parvient à confiner l'effet de fautes Byzantines avec un rayon constant. Nous avons de plus montr\'e que ce protocole fournit le rayon de confinement optimal pour ce probl\`eme.

Des travaux futurs sont nécessaires pour déterminer la qualité globale du mariage obtenu par notre protocole. Il est connu que, dans un scénario sans Byzantins, tout mariage maximal \cite{HH92j,MMPT09j} est à un facteur $2$ de l'optimal (le mariage maximum), mais il existe de meilleures solutions par rapport au facteur d'approximation \cite{MMPT11j}. Il serait intéressant d'étendre ces travaux en présence de Byzantins.

\vspace{-0.1cm}
\bibliographystyle{alpha}
\small{
\bibliography{biblio}
}

\end{document}